\documentclass[aps,prd,reprint,nofootinbib,superscriptaddress,floatfix]{revtex4-2}

\usepackage[T1]{fontenc}
\usepackage[utf8]{inputenc}
\usepackage{lmodern}
\usepackage{amsmath,amssymb,amsfonts,bm}
\usepackage{graphicx}
\usepackage{booktabs}
\usepackage{placeins}
\usepackage{microtype}
\usepackage[hidelinks]{hyperref}

\newcommand{\GeV}{\mathrm{GeV}}
\newcommand{\Mcal}{\mathcal M}
\newcommand{\Dcal}{\mathcal D}
\newcommand{\Tr}{\mathrm{Tr}}

\begin{document}

\title{Gottfried--Jackson polarization filters for holographic dilaton and transverse-traceless gluon responses in near-threshold $J/\psi$ photoproduction}
\author{Arkadiy I.\ Syamtomov}
\email{arkady.syamtomov@gmail.com}
\affiliation{Bogolyubov Institute for Theoretical Physics, National Academy of Sciences of Ukraine, Kyiv, Ukraine}

\begin{abstract}
Within the Hatta--Yang holographic graviton--dilaton amplitude for near-threshold
$J/\psi$ photoproduction, we identify exact dilaton-null spin-projection channels in
the Gottfried--Jackson basis.  The $m=0$ and $m=-\lambda_\gamma$ amplitudes receive
no dilaton contribution; for photon polarization normal to the reaction plane the
$m=0$ channel further isolates $A_g(t)+B_g(t)$ and requires nucleon spin transfer.
Standard beam-polarization spin-density-matrix elements then furnish
a TT-sensitive combination and a second bilinear containing linear
dilaton--TT interference.  Comparison with modern GlueX and CLAS12
cross sections shows substantial sensitivity to the shape of the
original holographic parameterization.  The polarization observables
therefore provide direct probes of the tensor structure of the
production amplitude.
\end{abstract}

\maketitle

\section{Introduction}

In a number of recent studies, near-threshold exclusive $J/\psi$ photoproduction has been used to constrain gluonic gravitational form factors (GFFs) and proton mass distributions~\cite{HattaYang2018,HattaRajanYang2019,MamoZahed2020,MamoZahed2022,Duran2023,GuoYuanZhao2025,PentchevChudakov2025,CLAS122026}.  The underlying motivation is the compact size of the $J/\psi$, which makes heavy quarkonium sensitive to gluonic fields in hadronic matter~\cite{Peskin1979,KharzeevSatzSyamtomovZinovjev1999}.  GlueX, $J/\psi$-007, and CLAS12 now provide differential data close to threshold~\cite{GlueX2023,Duran2023,CLAS122026}.  The extracted GFFs nevertheless remain dependent on the production
amplitude, and coupled-channel dynamics can contribute in the same
kinematic region~\cite{WinneyJPAC2023}.  The recent neutron measurement
provides an additional constraint on the underlying near-threshold
production mechanism~\cite{Neutron2026}.

Polarization provides information beyond that contained in the
unpolarized rate.  The density-matrix formalism for vector-meson photoproduction with polarized photons, including elementary $0^\pm$ exchange and naturality separation, was established long ago~\cite{SchillingSeybothWolf1970}.  Modern formulations and high-precision GlueX measurements use the same framework~\cite{Mathieu2018,GlueXRhoSDME2023}; double-polarization observables have also been proposed for near-threshold $J/\psi$ production as a discriminator of resonance mechanisms~\cite{WinneyEtAl2019}.  The aim of the present work is more specific.  We ask how this known polarization structure acts on the holographic graviton--dilaton amplitude of Hatta and Yang~\cite{HattaYang2018}, where the scalar exchange is tied to the renormalized gluonic scalar source and the non-scalar exchange to a $\Delta$-transverse, traceless projection of the gluon energy-momentum tensor (EMT).

This distinction is important at finite momentum transfer.  At finite momentum transfer, the Hatta--Yang TT graviton selects the
$\Delta$-transverse, traceless projection of the nonforward gluon EMT.
The complete local symmetric-traceless gluon matrix element additionally
contains the $D$ term, which drops out of this projection.  We therefore
distinguish throughout between the \emph{dilaton/scalar} and
\emph{TT/graviton} contributions.  The corresponding finite-$t$
projector comparison is given in Appendix~\ref{app:polarization}.

\section{Polarization projections}

We use the covariant production amplitude and nucleon vertices of Ref.~\cite{HattaYang2018}.  With photon momentum $q$, $J/\psi$ momentum $k$, photon polarization $\epsilon$, and $J/\psi$ polarization $\xi$, the amplitude can be written as
\begin{equation}
 \Mcal= X\,\Pi^{\mu\nu}\,\bar u(p')\Gamma^{\rm TT}_{\mu\nu}u(p)
 +Y\,\Pi^\mu{}_{\mu}\,\Gamma_{\phi}(t)\,\bar u(p')u(p).
 \label{eq:master}
\end{equation}
Here $X$ and $Y$ denote the graviton and dilaton normalizations, while
$\Pi^{\mu\nu}$ is the symmetric gauge-invariant tensor associated with the
upper $\gamma J/\psi$ transition vertex.  It may equivalently be written as
$\Pi^{\mu\nu}=F_\gamma^{(\mu|\alpha|}
F_{\psi\,\alpha}^{\ \ \nu)}$, so that
$\Pi^\mu{}_\mu=F_\gamma^{\mu\nu}F_{\psi\,\mu\nu}$.
In the Hatta--Yang amplitude the contraction with the TT nucleon vertex
selects the $A_g$ and $B_g$ form factors, whereas the trace
$\Pi^\mu{}_\mu$ multiplies $\Gamma_\phi$, which contains the scalar
combination of quark and gluon GFFs. The separate quark and gluon contributions depend on the renormalization prescription and scale~\cite{HattaRajanTanaka2018,Tanaka2019,BurkertRMP2023}.  The numerical parameter set of Ref.~\cite{HattaYang2018} is
therefore retained in its original model convention, providing a
well-defined realization of the graviton--dilaton amplitude for the
polarization analysis below.

We now choose the Gottfried--Jackson (GJ) frame.  The produced $J/\psi$ is at rest and the spin axis is the direction of the incident real photon.  The photon helicity is $\lambda=\pm1$, while $m=0,\pm1$ denotes the spin projection of the $J/\psi$ on the GJ axis.  Evaluating the trace of the production tensor gives
\begin{equation}
 \boxed{\Pi^\mu{}_{\mu;\lambda m}
 =-2(q\!\cdot\! k)\,\delta_{\lambda m}}.
 \label{eq:selection}
\end{equation}
A compact derivation and the phase convention are given in Appendix~\ref{app:polarization}.  Equation~\eqref{eq:selection} has the familiar elementary-scalar polarization pattern~\cite{SchillingSeybothWolf1970}; its significance here is the operator assignment in Eq.~\eqref{eq:master}.  The dilaton amplitude therefore satisfies
\begin{equation}
 \Mcal^{(\phi)}_{0,\lambda}=0,
 \qquad
 \Mcal^{(\phi)}_{-\lambda,\lambda}=0,
 \label{eq:nullchannels}
\end{equation}
whereas $\Mcal^{(\phi)}_{\lambda,\lambda}$ is in general nonzero.  Thus the $m=0$ and opposite-transverse-spin-projection channels are dilaton-null in the Hatta--Yang amplitude.  Additional production mechanisms can populate these channels, so their
measurement also provides sensitivity to contributions beyond the
leading graviton--dilaton amplitude.

Naturality provides a complementary classification.  Both the
$0^{++}$ scalar and $2^{++}$-type graviton exchanges have natural
parity, while Eq.~\eqref{eq:nullchannels} distinguishes them through
their different tensor couplings~\cite{SchillingSeybothWolf1970,Mathieu2018}.  The distinction follows instead from their different tensor couplings in Eq.~\eqref{eq:master}.

A second useful projection is obtained with a photon polarized normal to the reaction plane.  Taking the plane to be $xz$, $\bm\epsilon_\gamma=\hat{\bm y}$, and producing the GJ state $m=0$, the dilaton term vanishes and the TT contribution reduces to
\begin{equation}
 \boxed{\Mcal_{\perp\to0}
 =i\,\mathcal K(W,t)\,[A_g(t)+B_g(t)]
 \left(-r_z\sigma_x+r_x\sigma_z\right),}
 \label{eq:ABfilter}
\end{equation}
with
\begin{align}
 r_i&=\frac{p_i}{E+M}-\frac{p_i'}{E'+M},\\
 \mathcal K&=X\sqrt{(E+M)(E'+M)}\,
 \frac{q\!\cdot\!k}{2}(\bar P^0-\bar P^z).
 \label{eq:Kexplicit}
\end{align}
The result is independent of $D_g$ and $\bar C_g$ and is purely nucleon-spin transferring.  Its connection with the forward gluon angular momentum is obtained
through extrapolation to $t=0$, where
$J_g=\tfrac12[A_g(0)+B_g(0)]$.

\section{Comparison with existing cross sections}

To establish the numerical context for the polarization analysis, we
compare the same fixed-shape amplitude with the published unpolarized
proton data.  The GlueX measurement covers $8.2<E_\gamma<11.44~\GeV$ and reports both total and differential cross sections over the full physical $t$ range~\cite{GlueX2023}; the recent CLAS12 measurement provides an independent determination over the overlapping near-threshold region~\cite{CLAS122026}.  Since the overall photon--$J/\psi$ coupling is not fixed by the holographic construction, we multiply the $X=1$ cross section by one common normalization $\mathcal N$ for each of the two limiting choices $b=0$ and $b=1$
considered in Ref.~\cite{HattaYang2018}, leaving the $E_\gamma$ and $t$ dependence unchanged.

Using statistical and point-to-point systematic errors in quadrature, a common-normalization fit to the 18 GlueX and 10 CLAS12 total-cross-section points gives
\begin{equation}
 \chi^2/\nu=3.65\quad (b=0),\qquad
 \chi^2/\nu=5.29\quad (b=1),
 \label{eq:crossfit}
\end{equation}
with $\nu=27$.  GlueX quotes an additional fully correlated $19.5\%$ normalization uncertainty~\cite{GlueX2023,GlueXHEPData2023}; including it as a Gaussian nuisance parameter reduces these values only to $3.36$ and $4.86$, with normalization pulls of $-1.67\sigma$ and $-2.12\sigma$, respectively.  The CLAS12 data alone are compatible with the predicted energy
dependence after an overall rescaling, with
$\chi^2/\nu=0.38$ and $0.89$ for $b=0$ and $b=1$, respectively.
The combined GlueX and CLAS12 data display substantially greater
sensitivity to the fixed form-factor shapes.  This sensitivity is
consistent with the recent CLAS12 analysis, where allowing the GFF
shape parameters to vary gives $\chi^2/\mathrm{ndf}=1.05$ in a
combined GlueX, $J/\psi$-007, and CLAS12 holographic-QCD fit
~\cite{CLAS122026}.  The present comparison isolates this shape
dependence by retaining the original Hatta--Yang form factors and
varying only the common normalization.

\begin{figure}[t]
 \centering
 \includegraphics[width=\columnwidth]{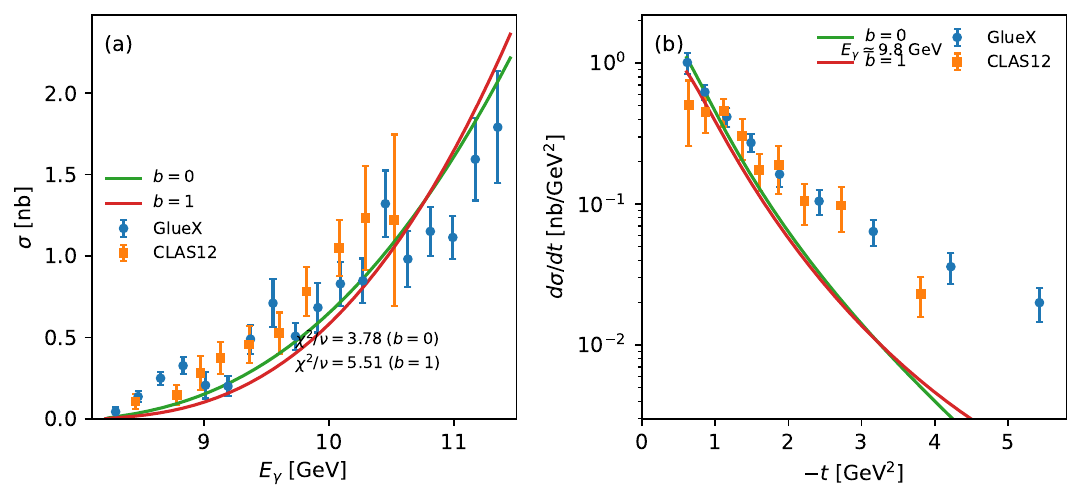}
 \caption{Comparison of the original Hatta--Yang amplitude with published proton data.  (a) GlueX~\cite{GlueX2023} and CLAS12~\cite{CLAS122026} total cross sections.  The $b=0$ and $b=1$ curves use one normalization fitted to all unshifted points; the quoted $\chi^2/\nu$ values correspond to this common-scale fit.  The fully correlated $19.5\%$ GlueX normalization uncertainty is
treated separately through the normalization nuisance parameter.  (b) Differential cross sections in the overlapping $E_\gamma\simeq9.7$--$9.9~\GeV$ region.  Panel (b) uses the normalizations determined from the total-cross-section
comparison in panel (a).}
 \label{fig:data}
\end{figure}

Figure~\ref{fig:data}(b) provides complementary information on the
$t$ dependence.  Around $E_\gamma\simeq9.8~\GeV$ the calculation
reproduces the rapid near-forward decrease, while the data favor a
slower falloff at moderate and large $-t$.  The complete comparison
at three energies is given in Appendix~\ref{app:numerics}.  Together with the structures observed
by GlueX and coupled-channel analyses of this energy region
~\cite{GlueX2023,WinneyJPAC2023}, this behavior motivates additional
dynamical contributions to the near-threshold amplitude.  We use the
holographic amplitude below to isolate the polarization consequences
of its graviton--dilaton tensor decomposition.

\section{Polarization observables and form-factor sensitivity}

Let $T_{m\lambda}$ denote the $2\times2$ matrix in nucleon-spin space for photon helicity $\lambda$ and GJ spin projection $m$.  It is convenient to write
\begin{equation}
 H_\lambda=T_{\lambda,\lambda},\qquad
 L_\lambda=T_{0,\lambda},\qquad
 F_\lambda=T_{-\lambda,\lambda}.
 \label{eq:blocks}
\end{equation}
Within Eq.~\eqref{eq:master}, $H_\lambda=G_\lambda+S_\lambda$ contains TT and dilaton pieces, while $L_\lambda$ and $F_\lambda$ contain TT contributions only.  Using the standard Schilling--Seyboth--Wolf convention~\cite{SchillingSeybothWolf1970,Mathieu2018}, and denoting by $(A|B)=\tfrac12\Tr(AB^\dagger)$ the nucleon-spin product, one finds
\begin{align}
 \frac{\Dcal}{2}
 \left(\mathrm{Re}\,\rho^1_{10}-\mathrm{Im}\,\rho^2_{10}\right)
 &=\mathrm{Re}(H_+|L_-),
 \label{eq:sensitive}\\
 \frac{\Dcal}{2}
 \left(\mathrm{Re}\,\rho^1_{10}+\mathrm{Im}\,\rho^2_{10}\right)
 &=\mathrm{Re}(F_-|L_+),
 \label{eq:control}
\end{align}
where $\Dcal$ is the unpolarized intensity.  These are standard SDME bilinears; their usefulness here follows from Eq.~\eqref{eq:nullchannels}.  Equation~\eqref{eq:control} is a TT-only control, while
\begin{equation}
 \mathrm{Re}(H_+|L_-)=
 \mathrm{Re}(G_+|L_-)+\mathrm{Re}(S_+|L_-)
 \label{eq:split}
\end{equation}
contains a term linear in the dilaton--TT interference.

The nine SDMEs accessible with an unpolarized target and a linearly
polarized beam provide independent constraints and null tests on the
photoproduction amplitude.  After parity is imposed, vector-meson
photoproduction on a spin-$1/2$ target contains 12 independent complex
amplitudes~\cite{PichowskySavkliTabakin1996}; target and recoil
polarization observables supply the complementary phase information
needed for a complete amplitude reconstruction.  The detailed counting
and the relation to the $J/\psi\to\ell^+\ell^-$ angular distribution
are summarized in Appendix~\ref{app:sdme}.

The unpolarized GJ element
\begin{equation}
 \rho^{0,\mathrm{GJ}}_{00}
 =\frac{\sum_\lambda\|L_\lambda\|^2}{\Dcal}
 \label{eq:rho00}
\end{equation}
has a dilaton-null numerator, although the denominator contains all amplitudes.  It therefore measures the TT-sensitive numerator relative to the full
unpolarized intensity, with scalar dependence entering through the
normalization $\Dcal$.

\begin{figure}[t]
 \centering
 \includegraphics[width=\columnwidth]{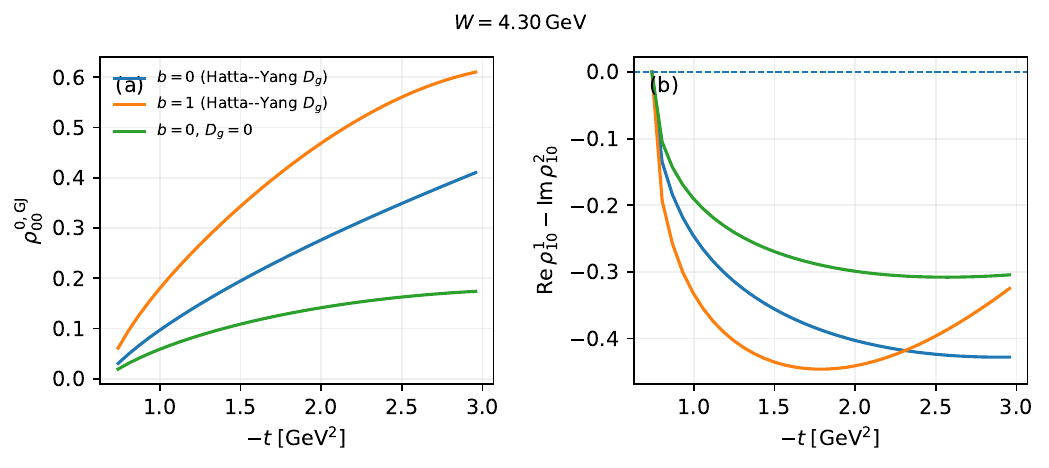}
 \caption{Polarization observables from the original Hatta--Yang
parameterization at $W=4.30~\GeV$.
(a) $\rho^{0,\mathrm{GJ}}_{00}$, whose numerator is dilaton-null.
(b) The SDME combination in Eq.~\eqref{eq:sensitive}.
The curves correspond to the limiting $b=0$ and $b=1$
mass-decomposition choices of Ref.~\cite{HattaYang2018}; the
additional $b=0$ curve shows the result obtained after setting
the gluon $D$ term to zero.}
 \label{fig:sdme}
\end{figure}

For the numerical illustration we adopt the original parameter set of
Ref.~\cite{HattaYang2018}, including $Y/X=-11/80$.  The overall
normalization determined from the cross-section comparison cancels from
the normalized SDMEs.  In the convention of Ref.~\cite{HattaYang2018},
$b=0$ and $b=1$ correspond to the limiting maximal- and
vanishing-anomaly cases, respectively.  With these inputs the predicted
cross section exhibits the characteristic enhancement of the $b=0$
case relative to $b=1$ near $W\simeq4.06~\GeV$.

Figure~\ref{fig:sdme} illustrates the resulting polarization dependence
at $W=4.30~\GeV$.  To expose the role of the gluon $D$ term, we also
show the $b=0$ result obtained after setting $D_g=0$.  The pronounced
change in the scalar-sensitive combination demonstrates its strong
dependence on $D_g$ within the original Hatta--Yang parameterization.

To examine how strongly this dependence is tied to the original
tripole ansatz, Appendix~\ref{app:numerics}
also considers the effect of replacing the gluon $A_g$ and $D_g$
form factors by the phenomenological parameterization of
Ref.~\cite{WangZengWang2023}, while retaining the remaining
Hatta--Yang inputs.  The resulting variation shows that the strong $D_g$ sensitivity
persists for this alternative form-factor shape.  This mixed
parameterization serves to illustrate the form-factor dependence;
a fully quantitative comparison would additionally require a common
fit and renormalization prescription.

These beam-SDME combinations can be extracted through the standard
dilepton angular analysis with a linearly polarized photon beam.  The $J/\psi\to e^+e^-$ or $\mu^+\mu^-$ decay angular distribution provides access to the spin-density matrix~\cite{H1Jpsi2006}, while linear photon polarization separates the $\rho^1$ and $\rho^2$ contributions~\cite{SchillingSeybothWolf1970,Mathieu2018}.  A complete reconstruction of the scalar--TT relative phase additionally requires target and/or recoil polarization.  GlueX has demonstrated precise linearly polarized SDME measurements for light vector mesons~\cite{GlueXRhoSDME2023}. In the recent CLAS12 near-threshold analysis the polarization
dependence enters through the acceptance systematic
~\cite{CLAS122026}.  A dedicated SDME measurement would therefore add
new polarization information to the published proton cross-section
data.

The selection rule in Eq.~\eqref{eq:selection} is a characteristic
prediction of the leading Hatta--Yang tensor structure.  Contributions
from higher-derivative bulk interactions, finite-coupling effects,
vector-meson dominance, open-charm rescattering, or additional
exchanges can populate the same polarization channels.  Measurements
of these channels therefore provide a direct way to quantify the
relative importance of such mechanisms.  Agreement with the predicted
null pattern would support the graviton--dilaton decomposition and
allow the $m=0$ and opposite-transverse-spin sectors to constrain the
TT component before the diagonal GJ block is used to determine the
additional scalar contribution.

\section{Summary}

We have applied the established polarized-vector-meson formalism to
the Hatta--Yang holographic graviton--dilaton amplitude for
near-threshold $J/\psi$ photoproduction.  In the GJ basis the dilaton
occupies the diagonal $m=\lambda_\gamma$ block, while the $m=0$ and
opposite-transverse-spin sectors provide exact dilaton-null
projections of this amplitude.  For photon polarization normal to the
reaction plane, the $m=0$ channel further isolates
$A_g(t)+B_g(t)$ with nucleon spin transfer.  Beam-polarization SDMEs
then provide experimentally accessible combinations that separately
probe the TT sector and its interference with the dilaton contribution.
The Hatta--Yang graviton selects a definite finite-$t$ TT projection
of the nonforward gluon EMT, making the resulting GJ polarization
pattern a direct probe of this tensor structure.  The comparison with
GlueX and CLAS12 cross sections shows substantial sensitivity to the
shape of the original 2018 parameterization and motivates refinements
of the reaction dynamics.  Polarization measurements add an
independent dimension to this program: the dilaton-null channels,
the $A_g+B_g$ projection, and the scalar--TT interference combinations
provide complementary constraints on the gluonic structure of the
near-threshold amplitude beyond those available from unpolarized
cross sections.

\section*{Data Availability}
The numerical values underlying the figures, the published GlueX and
CLAS12 cross-section data used in Fig.~\ref{fig:data}, and the scripts
used for the numerical evaluation are provided as ancillary files with
this arXiv submission.
This study uses published experimental data together with the
theoretical and numerical results presented here.

\clearpage
\onecolumngrid
\appendix

\section{Polarization identities and tensor projections}
\label{app:polarization}

\subsection{Gauge-invariant production tensor and GJ projection}
\label{app:gj}

We use $g^{\mu\nu}=\mathrm{diag}(1,-1,-1,-1)$.  The Hatta--Yang upper vertex can be written as
\begin{align}
\Pi^{\mu\nu}={}&q^{(\mu}k^{\nu)}\,\epsilon\!\cdot\!\xi^*
+\epsilon^{(\mu}\xi^{*\nu)}\,q\!\cdot\!k
-q^{(\mu}\xi^{*\nu)}\,k\!\cdot\!\epsilon
-k^{(\mu}\epsilon^{\nu)}\,q\!\cdot\!\xi^* .
\label{eq:PiApp}
\end{align}
Equivalently, $\Pi^{\mu\nu}=F_\gamma^{(\mu|\alpha|}F_{\psi\,\alpha}^{\ \ \nu)}$, with
$F_\gamma^{\mu\nu}=q^\mu\epsilon^\nu-q^\nu\epsilon^\mu$ and
$F_\psi^{\mu\nu}=k^\mu\xi^{*\nu}-k^\nu\xi^{*\mu}$.
Gauge invariance is then immediate, and the trace is
\begin{equation}
\Pi^\mu{}_{\mu}=2\left[(q\!\cdot\!k)(\epsilon\!\cdot\!\xi^*)
-(q\!\cdot\!\xi^*)(k\!\cdot\!\epsilon)\right].
\label{eq:PitraceApp}
\end{equation}

In the GJ frame a convenient longitudinal polarization vector is
\begin{equation}
\xi_0^\mu=\frac{M_\psi}{q\cdot k}q^\mu-\frac{k^\mu}{M_\psi},
\qquad k\cdot\xi_0=0,
\qquad \xi_0^2=-1,
\end{equation}
while for $m=\pm1$ we choose
$\epsilon_\lambda\cdot\xi_m^*=-\delta_{\lambda m}$ and
$q\cdot\xi_m^*=0$.  Substitution in Eq.~\eqref{eq:PitraceApp} directly gives the selection rule in Eq.~\eqref{eq:selection}.  In particular, for $m=0$ one uses
$q\cdot\xi_0=-(q\cdot k)/M_\psi$ and
$\epsilon\cdot\xi_0=-(\epsilon\cdot k)/M_\psi$, so the two terms cancel.
The same polarization pattern is familiar from elementary scalar exchange~\cite{SchillingSeybothWolf1970}; here its role is fixed by the dilaton/TT operator assignment in Eq.~\eqref{eq:master}.

\subsection{Perpendicular polarization and the $A_g+B_g$ projection}
\label{app:abfilter}

Take the reaction plane to be $xz$, the incoming photon along the GJ $z$ axis, and $\bm\epsilon=\hat{\bm y}$.  Since all hadron momenta lie in the reaction plane,
$\epsilon_y\cdot\bar P=\epsilon_y\cdot\Delta=\epsilon_y\cdot k=0$.
For $m=0$ the dilaton contribution vanishes, and contraction with the TT nucleon vertex of Ref.~\cite{HattaYang2018} gives
\begin{equation}
\Mcal_{\perp\to0}=X[A_g(t)+B_g(t)]\frac{(q\cdot k)(\bar P^0-\bar P^z)}{2}\,
\bar u(p')\gamma^y u(p).
\label{eq:ABdiracApp}
\end{equation}
For canonical Dirac spinors,
\begin{equation}
\bar u(p')\gamma^y u(p)=i\sqrt{(E'+M)(E+M)}
\left(-r_z\sigma_x+r_x\sigma_z\right),
\end{equation}
with $r_i=p_i/(E+M)-p_i'/(E'+M)$.  Combining these expressions yields Eqs.~\eqref{eq:ABfilter}--\eqref{eq:Kexplicit}.

\subsection{Finite-$t$ gluon EMT and the TT projection}
\label{app:emt}

For the standard nucleon GFF decomposition,
\begin{align}
\langle p'|T_g^{\mu\nu}|p\rangle=\bar u'\Big[&A_g\gamma^{(\mu}\bar P^{\nu)}
+B_g\frac{\bar P^{(\mu}i\sigma^{\nu)\rho}\Delta_\rho}{2M}
+D_g\frac{\Delta^\mu\Delta^\nu-g^{\mu\nu}t}{4M}
+\bar C_g M g^{\mu\nu}\Big]u .
\label{eq:GFFApp}
\end{align}
A Lorentz-traceless chromoelectric projection and the $\Delta$-transverse, traceless graviton projection select different finite-$t$ combinations.  For the aligned chromoelectric direction one finds
\begin{equation}
\widehat R_E^{(2)}=A_g+\eta B_g-\frac{\eta}{3}D_g,
\qquad \eta=\frac{t}{4M^2},
\end{equation}
whereas the transverse TT graviton projection gives
\begin{equation}
\widehat R_{hT}=A_g+\eta B_g.
\end{equation}
Thus the $D_g$ contribution drops out of the Hatta--Yang TT projector while remaining part of the full nonforward symmetric-traceless gluon matrix element.

\section{SDME conventions and amplitude counting}
\label{app:sdme}

Let $T_{m\lambda}$ be the $2\times2$ nucleon-spin amplitude for GJ spin projection $m$ and photon helicity $\lambda$, and define
\begin{equation}
(A|B)=\frac12\Tr(AB^\dagger),
\qquad
\Dcal=\sum_{m,\lambda}(T_{m\lambda}|T_{m\lambda}).
\end{equation}
The real-photon Schilling--Seyboth--Wolf density matrices are
\begin{align}
\rho^0_{mm'}&=\frac{(T_{m+}|T_{m'+})+(T_{m-}|T_{m'-})}{\Dcal},\\
\rho^1_{mm'}&=\frac{(T_{m+}|T_{m'-})+(T_{m-}|T_{m'+})}{\Dcal},\\
\rho^2_{mm'}&=\frac{-i[(T_{m+}|T_{m'-})-(T_{m-}|T_{m'+})]}{\Dcal}.
\label{eq:rhoDefsApp}
\end{align}
For a linearly polarized beam,
\begin{equation}
\rho(\Phi)=\rho^0-P_\gamma\left[\rho^1\cos2\Phi+\rho^2\sin2\Phi\right],
\label{eq:rhoPhiApp}
\end{equation}
where $\Phi$ is the angle between the photon-polarization and production planes.  Contracting $\rho_{mm'}(\Phi)$ with the standard vector-to-dilepton decay matrices gives the measured $J/\psi\to\ell^+\ell^-$ distribution; the standard angular functions follow Refs.~\cite{SchillingSeybothWolf1970,Mathieu2018} and have been used in $J/\psi$ analyses such as Ref.~\cite{H1Jpsi2006}.
With the blocks defined in Eq.~\eqref{eq:blocks}, Eq.~\eqref{eq:rhoDefsApp} directly yields Eqs.~\eqref{eq:sensitive} and \eqref{eq:control}.

Before symmetry reduction, real-photon vector-meson production on a spin-$1/2$ target contains $2\times3\times2\times2=24$ complex amplitudes.  Parity leaves 12 independent complex amplitudes~\cite{PichowskySavkliTabakin1996}, corresponding to 23 physical real parameters after removal of one overall phase.  A linearly polarized beam on an unpolarized target supplies nine independent vector-meson SDMEs; target and recoil polarization observables provide the complementary phase information needed for a complete amplitude reconstruction.

\section{Numerical details and additional comparisons}
\label{app:numerics}

For the numerical results we use the parameters of Ref.~\cite{HattaYang2018}:
$M=0.94~\GeV$, $M_\psi=3.10~\GeV$, $A_g(0)=0.43$,
$A_q(0)=0.57$, $\Lambda^2=0.71~\GeV^2$, $B_g=B_q=0$,
$n_f=3$, $\gamma_m=0.20$, $g/(2\beta)=-2.20$, and
$Y/X=-11/80$, together with the form factors specified there.
The parameter $b$ controls the forward scalar mass decomposition;
$b=0$ and $b=1$ are the limiting maximal- and vanishing-anomaly cases.

\subsection{Energy dependence and cross-section comparison}
\label{app:crosssections}

\begin{figure}[!ht]
\centering
\includegraphics[width=0.62\textwidth]{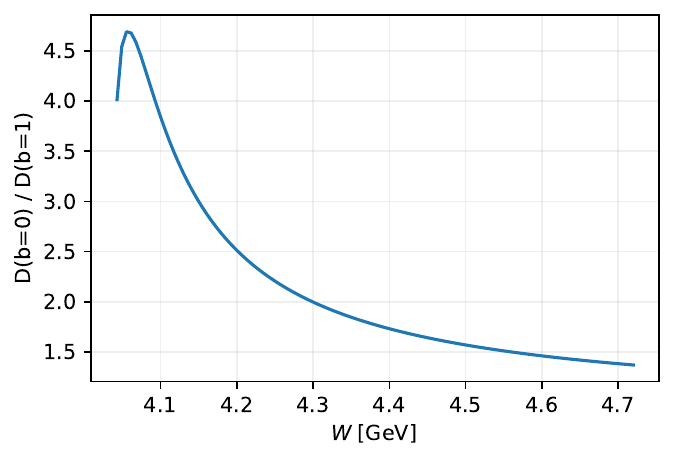}
\caption{Ratio of the unpolarized Hatta--Yang intensities for the $b=0$ and $b=1$ parameter choices in forward kinematics.  The characteristic enhancement near $W\simeq4.06~\GeV$ follows the behavior obtained in Ref.~\cite{HattaYang2018}.}
\label{fig:validation}
\end{figure}

Figure~\ref{fig:validation} shows the forward ratio of unpolarized intensities for the two limiting choices.  The characteristic enhancement near $W\simeq4.06~\GeV$ follows Ref.~\cite{HattaYang2018}.

\FloatBarrier

For the published GlueX and CLAS12 total cross sections, the fixed-shape theory can be written as
\begin{equation}
\sigma_i^{\rm th}=\mathcal N f_i^{(b)}.
\end{equation}
The direct common-normalization fit minimizes
\begin{equation}
\chi^2=\sum_i\frac{[\sigma_i^{\rm exp}-\mathcal N f_i^{(b)}]^2}{\delta_i^2},
\end{equation}
and the correlated GlueX normalization uncertainty can be included through
\begin{align}
\chi^2={}&\sum_{i\in G}\frac{[\sigma_i^{\rm exp}-\mathcal N(1+\delta_G)f_i^{(b)}]^2}{\delta_i^2}
+\sum_{i\in C}\frac{[\sigma_i^{\rm exp}-\mathcal N f_i^{(b)}]^2}{\delta_i^2}
+\left(\frac{\delta_G}{0.195}\right)^2 .
\end{align}
The resulting values are summarized in Table~\ref{tab:crossfit}.

\begin{table}[!ht]
\caption{Normalization fits of the original Hatta--Yang $b=0,1$ parameter choices to the published total cross sections.  The last column gives the fitted GlueX normalization nuisance in units of its quoted $19.5\%$ correlated uncertainty.}
\label{tab:crossfit}
\centering
\begin{ruledtabular}
\begin{tabular}{ccccc}
$b$ & data set & $\chi^2$ & $\nu$ & $\chi^2/\nu$ \quad $\delta_G/0.195$\\
\hline
0 & CLAS12 only & 3.42 & 9 & 0.38 \quad --\\
1 & CLAS12 only & 8.04 & 9 & 0.89 \quad --\\
0 & GlueX+CLAS12 & 98.46 & 27 & 3.65 \quad --\\
1 & GlueX+CLAS12 & 142.95 & 27 & 5.29 \quad --\\
0 & GlueX+CLAS12 + norm. nuisance & 87.30 & 26 & 3.36 \quad -1.67\\
1 & GlueX+CLAS12 + norm. nuisance & 126.39 & 26 & 4.86 \quad -2.12\\
\end{tabular}
\end{ruledtabular}
\end{table}
\FloatBarrier

\begin{figure}[!ht]
\centering
\includegraphics[width=0.94\textwidth]{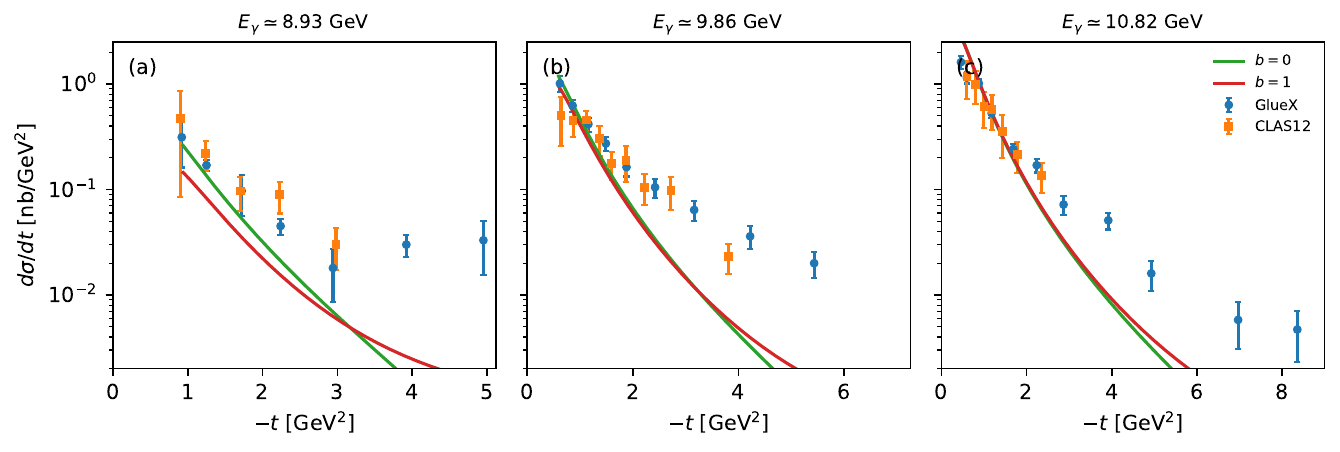}
\caption{Differential cross sections using the common normalizations determined from the total-cross-section comparison.  GlueX data are from Refs.~\cite{GlueX2023,GlueXHEPData2023} and CLAS12 data from Ref.~\cite{CLAS122026}.  The curves are evaluated at the representative energy indicated in each panel.  Statistical and point-to-point systematic uncertainties are combined in quadrature, while the correlated GlueX normalization uncertainty is treated separately.}
\label{fig:datafull}
\end{figure}

The differential measurements probe the momentum-transfer dependence with the normalization fixed by the total-cross-section comparison.  Figure~\ref{fig:datafull} shows three representative energy intervals; the data favor a slower falloff at moderate and large $-t$ than the original fixed-shape parameterization.

\FloatBarrier

The $J/\psi$-007 measurement~\cite{Duran2023} provides complementary two-dimensional differential information and can be incorporated in a future common multidimensional analysis with GlueX and CLAS12.

\subsection{Scalar--TT interference and gluon-GFF sensitivity}
\label{app:gff}

\begin{figure}[!ht]
\centering
\includegraphics[width=0.88\textwidth]{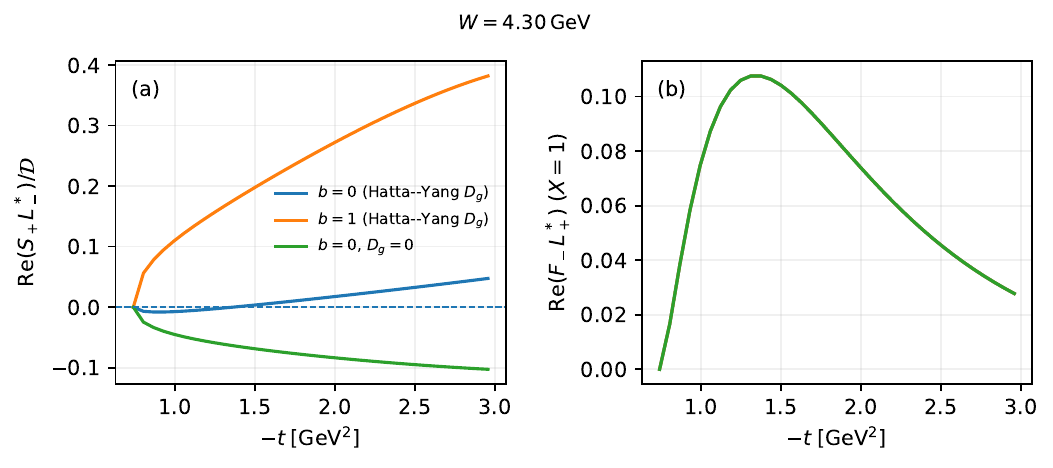}
\caption{At $W=4.30~\GeV$: (a) normalized dilaton--TT interference; (b) the unnormalized TT-sensitive numerator for $X=1$.  The coincident curves in panel (b) display directly the dilaton independence of this combination.}
\label{fig:S1}
\end{figure}

The additional polarization combinations are shown in Fig.~\ref{fig:S1}.  The coincident TT-sensitive numerator displays directly its dilaton independence.

\FloatBarrier

\begin{figure}[!ht]
\centering
\includegraphics[width=0.55\textwidth]{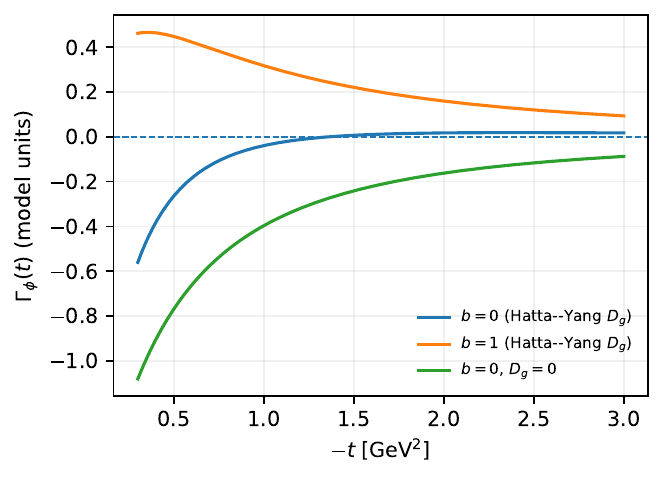}
\caption{Scalar coefficient $\Gamma_\phi(t)$ in the original Hatta--Yang parameterization.  The $b=0$ choice has a model-specific zero near $t\simeq-1.36~\GeV^2$, illustrating the nonmonotonic relation between the forward mass-decomposition parameter and the finite-$t$ scalar production amplitude.}
\label{fig:S2}
\end{figure}

The finite-$t$ scalar coefficient is shown in Fig.~\ref{fig:S2}; its zero for the $b=0$ choice illustrates the nonmonotonic relation between the forward mass decomposition and the scalar production amplitude.

\FloatBarrier

\begin{figure}[!ht]
\centering
\includegraphics[width=0.68\textwidth]{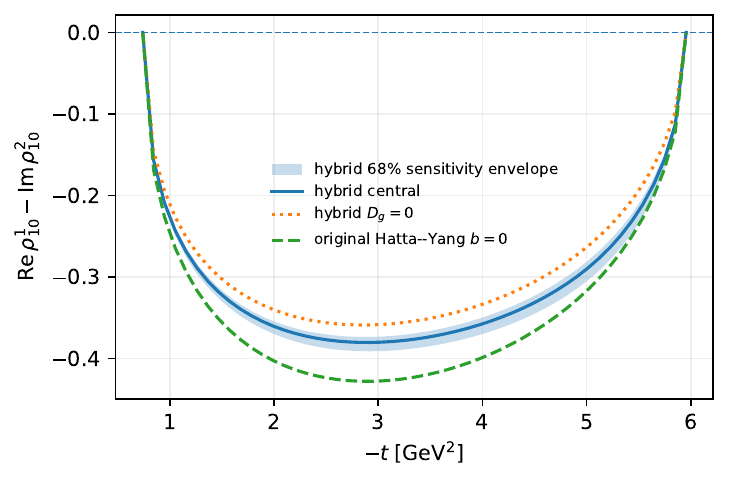}
\caption{Sensitivity to the gluon-GFF parameterization at $W=4.30~\GeV$.  The band shows the central $68\%$ variation obtained by sampling the quoted uncertainties of $m_g$, $D_g(0)$, and $d_g$ from Ref.~\cite{WangZengWang2023}, with the remaining Hatta--Yang inputs held fixed.}
\label{fig:modern}
\end{figure}

To examine the dependence on the gluon form-factor shape, we also use the phenomenological parameterization of Wang, Zeng, and Wang~\cite{WangZengWang2023},
\begin{equation}
A_g(t)=\frac{0.414}{(1-t/m_g^2)^2},\qquad
D_g(t)=\frac{D_0}{(1-t/d_g^2)^3},
\end{equation}
with independent Gaussian sampling of $m_g=1.51\pm0.10~\GeV$, $D_0=-1.97\pm0.25$, and $d_g=0.86\pm0.07~\GeV$, while the remaining Hatta--Yang inputs are kept fixed.  Figure~\ref{fig:modern} shows that the strong $D_g$ sensitivity persists for this alternative negative-$D_g$ shape.  Since the parameter covariance matrix is not used here, the band represents the induced parameter variation rather than a correlated confidence region.

\FloatBarrier

Lattice-QCD determinations likewise provide direct constraints on the gluon $D$ term~\cite{Pefkou2022,Hackett2024}, reinforcing the importance of retaining its finite-$t$ dependence.

\bibliography{references}
\end{document}